\documentclass{IEEEtran}
\usepackage{cite}
\usepackage{amsmath,amssymb,amsfonts,theorem,color}
\usepackage{algorithmic}
\usepackage{graphicx}
\usepackage{textcomp}

\usepackage{mathtools} 
\usepackage{graphicx,epsfig,psfrag}
\usepackage{subfigure}
\usepackage{tikz}
\usepackage{float,verbatim}
\usepackage{cite}
\usepackage{dsfont}
\usepackage{relsize}
\DeclareMathOperator{\col}{col}
\DeclareMathOperator{\im}{im}
\DeclareMathOperator{\diag}{diag}

\newcommand{\R}{\ensuremath{\mathbb R}}
\newcommand{\norm}[1]{\ensuremath{\left\| #1 \right\|}}

\usepackage[normalem]{ulem}

\usepackage{enumerate}
\usepackage[normalem]{ulem}

\newcommand\xqed[1]{%
	\leavevmode\unskip\penalty9999 \hbox{}\nobreak\hfill
	\quad\hbox{#1}}

\usepackage[prependcaption,colorinlistoftodos]{todonotes}

\newcommand{\calK}{\ensuremath{\mathcal{K}}}

\newcommand{\BP}{\noindent{\bf Proof. }}
\newcommand{\EP}{\hspace*{\fill} $\blacksquare$\smallskip\noindent}
\newcommand{\bse}{\begin{subequations}}
	\newcommand{\ese}{\end{subequations}}
\def\be{\begin{equation}}
\def\ee{\end{equation}}

\newcommand{\bbm}{\begin{bmatrix}}
	\newcommand{\ebm}{\end{bmatrix}}

\newcommand{\Fdes}{\mathcal{F}_{\rm des}}
\newcommand{\Fadm}{\mathcal{F}_{\rm att}^{\rm data\,}}
\newcommand{\FadmM}{\mathcal{F}_{\rm att}^{\rm model\,}}

{\theorembodyfont{\upshape}\newtheorem{theorem}{Theorem}}
{\theorembodyfont{\upshape}\newtheorem{lemma}[theorem]{Lemma}}
{\theorembodyfont{\upshape}}
{\theorembodyfont{\upshape}\newtheorem{corollary}[theorem]{Corollary}}
{\theorembodyfont{\upshape}\newtheorem{definition}[theorem]{Definition}}
{\theorembodyfont{\upshape}}
{\theorembodyfont{\upshape}\newtheorem{remark}[theorem]{Remark}}
{\theorembodyfont{\upshape}\newtheorem{problem}{Problem}}
{\theorembodyfont{\upshape}\newtheorem{example}[theorem]{Example}}
{\theorembodyfont{\upshape}\newtheorem{assumption}{Assumption}}

\begin{document}
\title{{{A Versatile Framework for} Data-Driven Control of\\ Nonlinear Systems}}
\author{Nima Monshizadeh, Claudio De Persis, and Pietro Tesi
\thanks{N. Monshizadeh and C. De Persis are with the Engineering and Technology Institute, University of Groningen, 9747AG, The Netherlands (e-mail: n.monshizadeh@rug.nl, c.de.persis@rug.nl).}
\thanks{ P. Tesi is with DINFO, University of Florence, 50139 Florence, Italy (e-mail: pietro.tesi@unifi.it). His work was supported by the EU under the Italian National Recovery and Resilience Plan (NRRP) of NextGenerationEU, partnership on “Telecommunications of the Future” (PE00000001 - program “RESTART”).}
}

\maketitle

\begin{abstract}
This note aims to provide a systematic {investigation} of direct data-driven control, enriching the existing literature not by adding another isolated result, but rather by offering a {unifying, versatile, and broad framework that enables the generation of novel results in this domain. We formulate the nonlinear design problem from a high-level perspective as a set of desired controlled systems and propose systematic procedures to synthesize data-driven control algorithms that meet the specified design requirements. Various examples are presented to demonstrate the applicability of the proposed approach and its ability to derive new insights and results, illustrating the novel contributions enabled by the framework.}
\end{abstract}

\section{Introduction}\label{sec:introduction}
The ability to harness data to directly influence system design and control strategies represents a significant advancement in control systems research. Conventional  methodologies often rely on explicit system identification followed by controller design, a process that may not only be cumbersome but also less effective for complex systems. In contrast, data-driven control offers a promising avenue by enabling the design of controllers directly from raw data, without the intermediary step of system modeling. This approach has become increasingly attractive with advancements and popularization of machine learning algorithms, which have spread through various branches of science and engineering, including automatic control.

The expanding interest in data-driven methodologies is fuelled by their potential to circumvent the limitations of model-based techniques, particularly in the realm of nonlinear systems where a priori given mathematical {models} may not be available. {Nonlinear systems are characterized by their rich dynamics, which are not amenable to linear analysis and require a more involved approach to control design.}
As such, the field has seen diverse data-guided strategies being proposed and tested, ranging from  virtual reference control \cite{campi2002virtual}, kernel-based design \cite{tanaskovic2017data}, intelligent PID \cite{fliess2013model} and sampled-data model-free control \cite{tabuada2017data,fraile2020data} to {nonlinear} data-enabled model predictive control \cite{huang2023robust}, {\cite{lazar2023basis}, \cite[Sec. 4]{berberich2024overview}}, dynamic mode decomposition \cite{proctor2016dynamic}, Koopman design \cite{korda2018linear}, {first-order approximation} \cite[Subsection V.B]{de2019formulas} and {Taylor-expansion \cite{guo2022data,cheah2022robust}}.

The current work belongs to the category of direct data-driven control design aiming to  reduce the task of controller design to programs that are stated in terms of data collected from the system.  Due to its inherent complexity, the landscape of direct data-driven control for nonlinear systems is fragmented with solutions that often target specific problem settings or system types. Examples include stabilization of bilinear systems \cite{bisoffi2020data,yuan2021data}, polynomial systems \cite{dai2020semi, guo2021data}, rational systems \cite{strasser2021data}, and flat systems \cite{alsalti2021data}. Such specialization is valuable as it provides concrete solutions to the control problems under investigation.
{Nonetheless, within the class of approaches that yield one-shot, data-dependent matrix inequalities for controller synthesis \cite{de2023learning-rev}, there remains a need for} a coherent framework that can address a broader spectrum of {nonlinear} control challenges. While the focus on particular instances or system models has driven substantial contributions, a unified approach could  illuminate the underlying principles, fostering even greater advancements in the field.

Recognizing the need and the benefits for such a unified approach, this note introduces a {systematic} framework for the direct data-driven design of nonlinear control systems. 
We begin by embedding the design specifications in a set of desired closed-loop systems. Subsequently, we establish systematic procedures to obtain the control gain from the system data such that the design specifications are enforced. The controller parametrization in \cite{de2019formulas}, \cite{de2023learning}, serves as an inspiring  element of the {data-driven} procedures laid out in this manuscript. {The proposed approach distinguishes itself through its \textit{unifying}, \textit{versatile}, and \textit{innovative} aspects, aiming not only to integrate several existing results into a cohesive framework but also to extend their applicability and enable the generation of new results in the domain of nonlinear direct data-driven control.} Furthermore, the approach provides a systematic treatment of deriving control algorithms from data. This systematic procedure ensures that diverse objectives, such as stabilization, performance, {and input-output properties can be seamlessly adapted and integrated into the design. Several cases are presented to highlight the versatility of the framework in systematically deriving new insights and results, showcasing its ability to enable novel contributions.}

The structure of the manuscript is as follows. The structure of the manuscript is as follows: The problem of interest is formulated in Section \ref{s:prob}, the solution is provided in Subsection \ref{ss:sol}, {and solvability conditions for the nonlinear control problem are discussed} in Subsection \ref{ss:nec}.
Extensions of the results to open systems are provided in Subsection \ref{ss:open}. Finally, the manuscript closes with conclusions in Section \ref{s:conc}.

	\vspace*{-1mm}
	\section{Problem setup}\label{s:prob}
	
	Consider a nonlinear system of the form
	$$
	x^+= f(x)+Bu, 
	$$
	where $x\in \R^n$, $f:\R^n \rightarrow \R^n$ {is locally Lipschitz}, and $B\in \R^{n\times m}$. The notation $x^+$ denotes either the state variables shifted in time or the time derivative of the states depending on the setting (i.e. continuous or discrete time). {We restrict our attention to nonlinear systems that are linear in the control input.} While the map $f$ and the input matrix $B$ are assumed to be unknown, we assume that there exists a \textit{known} library of nonlinear functions, denoted by $Z:\R^n \rightarrow \R^s$, such that $f$ can be written as $AZ(x)$ for some {unknown} matrix $A\in \R^{n\times s}$. Hence, the nonlinear system is represented by
	\be\label{e:model}
	x^+=AZ(x)+Bu.
	\ee
	Note that we allow $Z(\cdot)$ to be as exhaustive as necessary and potentially include functions that do not appear in $f$. {To avoid redundancy in the representation, 
		we assume that the matrix $B$ has full column rank.

	Under a nonlinear state feedback protocol $u=KZ(x)$, $K\in \R^{m\times s}$,  the closed-loop system admits the form
	$
	x^+= F_K Z(x)
	$
	with $F_K:=A+BK$, $F_K\in \R^{n\times s}$.
	The aim of the nonlinear controller is to impose some desired stability/performance properties on the closed-loop system. Rather than fixing a particular choice of a closed-loop, we consider a set of desired closed-loop dynamics. A set of desired closed-loop systems can be parametrized as
	\be\label{CLdes}
	x^+= F^\star Z(x), \qquad F^\star\in \mathcal{F}_{\rm des},
	\ee
	where any matrix $F^\star\in\mathcal{F}_{\rm des}$ is such that the closed-loop vector field $F^\star Z(x)$ meets the design criteria. The next example illustrates this parametrization. 
	
	\begin{example}\label{ex:fdes}
		First, consider the case of a linear discrete-time system, namely $Z(x)=x$. Suppose that the design criterion is to geometrically stabilize the closed-loop system with a specified rate of decay. 
		This gives rise to
		\[
		\Fdes=\{  F :  \exists P>0, \quad \rho F^T P F- P < 0\},
		\]
		where the parameter $\rho>1$ can be chosen to control the decay rate of the solutions.
		In case of nonlinear systems, with the same design criterion, the set $\Fdes$ modifies to
		\[
		\Fdes=\{  F :  \exists V\in \mathcal{V}, \quad \rho  V(FZ(x))) \leq V(x),  \;\forall x\in \R^n\},
		\]
		where $\mathcal{V}$ denotes the set of Lyapunov functions satisfying $
		\alpha \norm{x}^2\leq V(x)\leq \beta \norm{x}^2
		$
		for some $\alpha, \beta>0$. Restricting to quadratic Lyapunov functions, i.e. $V(x)=x^TPx$, the latter can be written more explicitly as
		\begin{align*}
		\Fdes=\{  F :\, &\exists P>0, \\
		&  \rho Z(x)^T F^TP F Z(x)- x^TPx<0, \forall x\in \R^n\}.
		\end{align*}
			Clearly, in the above examples, any $F^\star\in \Fdes$ is such that the closed-loop system \eqref{CLdes} meets the design criterion. 
			{We also note that the Lyapunov dissipation inequalities in the presented examples of $\Fdes$ can be relaxed to a neighborhood around the equilibrium to capture \textit{local} rather than \textit{global} stabilization requirements.}
		{\xqed{$\square$}}
	\end{example}
	The central problem of this note is formulated as follows. 
	
	\begin{problem}\label{prob}
		Given $\Fdes$, design a state feedback protocol $u=KZ(x)$ such that the resulting closed-loop system belongs to the set of desired systems in \eqref{CLdes}. 
		Equivalently, find $K$ such that  $A+BK \in \Fdes$.
	\end{problem}

	{We} take a data-driven approach towards this problem. We collect input-state data from the system and store them in matrices $U_0\in \R^{m\times N}$, $Z_0\in\R^{s\times N}$, and $X_1\in \R^{n\times N}$ satisfying
\be\label{data}
	X_1= AZ_0+BU_0.
	\ee
	The equality \eqref{data} is consistent with \eqref{e:model}; specifically, {the columns of} \( U_0 \) corresponds to the input samples, {those of} \( Z_0 \) corresponds to the library \( Z(\cdot) \) evaluated at the state samples, and {the columns of} \( X_1 \) {are the state samples shifted in time in discrete time, or the time derivative of the state in continuous time.}
	{As will be observed, access to data samples compensates for the lack of knowledge in the system dynamics, enabling a data-driven solution to Problem 1.}

	\section{Main results}
	
	\subsection{Solutions to Problem \ref{prob}}\label{ss:sol}
	
	{First, we emphasize that} not any desired closed-loop system is attainable. Noting \eqref{e:model}, a closed-loop system $x^+=F Z(x)$ is attainable  if and only if there exists $K$ such that $A+BK=F$. 
	Hence, the set of attainable, not necessarily desired, closed-loop systems can be parametrized by the following set:
	\be\label{Fadm-model}
	\FadmM:= \{F:  \im (A-F) \subseteq \im B \}.
	\ee

	We note that working with either the model or the data does not affect the choice or formulation of $\Fdes$. On the other hand,  since we aim to use the collected data as a proxy for the model, we require a counterpart of \eqref{Fadm-model} in terms of data.
	To this end, we define
	\be\label{Fadm}
	\Fadm:=\left \{F:  \im \bbm F \\ I_s \ebm \subseteq \im \bbm X_1 \\ Z_0 \ebm \right \}.
	\ee
	The following lemma states several properties of this set and its relation to the set of attainable closed-loop models $\FadmM$. 
	\begin{lemma}\label{l:prop}
		The set $\Fadm$ has the following properties:
		\begin{enumerate}[(i)]
		\item The set $\Fadm$ is nonempty if and only if $Z_0$ has full row rank. 
		\smallskip{}
		\item $\Fadm \subseteq \FadmM$.
		\smallskip{}
		\item 
		$\Fadm = \FadmM$ if and only if the data matrix $\bbm Z_0 \\ U_0 \ebm$ has full row rank.
		\end{enumerate}
	\end{lemma}
	
	\BP
	Item (i): If $\Fadm$ is nonempty, then $\im I_s \subseteq \im Z_0$ and thus $Z_0$ has full row rank.
	Conversely, if  $Z_0$ has full row rank, then $X_1Z_0^+\in \Fadm$ for any right inverse $Z_0^+$ of $Z_0$.   \\Item (ii):
	Observe that the set $\FadmM$ in \eqref{Fadm-model} can be equivalently written as 
	\be\label{nec-1}
	\FadmM=\left \{F: \im \bbm F \\ I_s \ebm \subseteq \im {\bbm A & B \\I_s & 0 \ebm} \right\}. 
	\ee
	The subspace inclusion $\Fadm \subseteq \FadmM$ then holds due to 
	\be\label{nec-2}
	\im \bbm X_1 \\ Z_0 \ebm=\im {\bbm A & B \\I_s & 0 \ebm} \bbm Z_0 \\ U_0 \ebm \subseteq 
	\im {\bbm A & B \\I_s & 0 \ebm},
	\ee
	where we used \eqref{data} to write the first equality.\\
	Item (iii): {{
		The inclusion in \eqref{nec-2} can be replaced by equality if and only if the matrix $\left[\begin{smallmatrix} Z_0 \\ U_0 \end{smallmatrix}\right]$ has full row rank, {where the ``only if" part follows from the fact that 
			the matrix $B$ and thus$\left[\begin{smallmatrix}
			A & B \\I_s & 0 \end{smallmatrix}\right]$ has full column rank.}} 
		{To conclude the proof, it remains to show that 
		\[
		\FadmM = \Fadm \Longleftrightarrow 
		\im \bbm X_1 \\ Z_0 \ebm=  
		\im {\bbm A & B \\I_s & 0 \ebm}
		\]
The `$\Leftarrow$' direction trivially follows from \eqref{Fadm} and \eqref{nec-1}. To prove the other direction, suppose $\FadmM = \Fadm $. 
Noting \eqref{nec-2}, it suffices to show that $\im [ \begin{smallmatrix} A & B \\I_s & 0 \end{smallmatrix}]\subseteq \im [\begin{smallmatrix} X_1 \\ Z_0 \end{smallmatrix}]$. Observe that for any $\Theta\in\R^{m\times s}$, the matrix 
$F:= A+ B\Theta$ belongs to $\FadmM$ and thus to $\Fadm$.
Hence, for any $\Theta\in\R^{m\times s}$, there exists a solution $\Xi$ to the following equation:
\[
\bbm A+B\Theta \\  I_s\ebm= \bbm X_1 \\ Z_0 \ebm \Xi.  
\]
Now, we select two choices $\Theta=0$ and $\Theta= \bar\Theta$, for  some 
$\bar \Theta\in \R^{m\times s}$ having full row rank.  Then, there exist $\Xi$ and $\bar\Xi$ satisfying 
\[
\bbm A & B \\I_s & 0 \ebm \bbm I_s  & I_s\\ 0  & \bar\Theta \ebm = \bbm X_1 \\ Z_0 \ebm \bbm \Xi & \bar \Xi \ebm.   
\]
By using the fact that  $[\begin{smallmatrix} I_s  & I_s\\ 0  & \bar \Theta \end{smallmatrix}]$ has full row rank, we  conclude that
$\im [ \begin{smallmatrix} A & B \\I_s & 0 \end{smallmatrix}]\subseteq \im [\begin{smallmatrix} X_1 \\ Z_0 \end{smallmatrix}]$, which completes the proof.}

\EP
	
	Observe that Problem \ref{prob} is solvable if and only if $\FadmM \cap \Fdes\ne \emptyset$; namely,  the desired set $\Fdes$ and the attainable set  $\FadmM$ should at least share a common element. Motivated by the fact that we work directly with the data rather than the model, we replace $\FadmM$ by its data-based {subset} $\Fadm$ in the latter solvability condition, which results in the following assumption:
	
	\begin{assumption}\label{a:intersect}
		It holds that $\Fadm \cap \Fdes \ne \emptyset$.
	\end{assumption}
	
	This assumption is sufficient for solvability of Problem \ref{prob} since $\Fadm \subseteq \FadmM$ by Lemma \ref{l:prop}.  We will discuss ``necessity" of this assumption in Subsection \ref{ss:nec}.

		We have now the following result:

	\begin{theorem}\label{t:main}
		Let Assumption \ref{a:intersect} hold and 
		define
		\be\label{calK}
		\mathcal{K}:=\Big\{ K : \im \bbm F^\star \\ I_s \\ K  \ebm  \subseteq \im \bbm X_1 \\ Z_0 \\ U_0 \ebm, \; {F^\star \in \Fdes} \Big\}. 
		\ee
		Then,  the set $\mathcal{K}$ is nonempty. Moreover, Problem \ref{prob} is solvable by the state feedback $u=KZ(x)$ for any $K\in \mathcal{K}$. 
	\end{theorem}

	\BP
	{The fact that the set $\calK$ is nonempty follows from Assumption 1 and \eqref{Fadm}.} 
	 Now, suppose $K\in \calK$. Then, we find that
	\be\label{pr:G}
	\bbm F^\star \\ I_s \\ K  \ebm  =\bbm X_1 \\ Z_0 \\ U_0 \ebm G,
	\ee
	for some matrices $G\in \R^{N\times s}$ {and $F^\star\in \Fdes$}.
	We have
	\[
	A+BK= \bbm A & B \ebm \bbm I \\ K \ebm = \bbm A & B \ebm \bbm Z_0\\ U_0 \ebm G = X_1G =F^\star,
	\]
	where the second and last equality follow from \eqref{pr:G}, and the third one from \eqref{data}. As $F^\star \in\Fdes$, we conclude that $(A+BK)\in \Fdes$ and the controller $KZ(x)$ solves Problem \ref{prob}. 
	\EP
	
	Theorem \ref{t:main} provides a {systematic} procedure for solving Problem \ref{prob} using data:

	\begin{enumerate}
		\item Parametrize $\Fdes$ based on the design objective. 
		\item Find $F^\star\in \Fdes \cap \Fadm$. Namely, find $F^\star$ such that\footnote{Such $F^\star$ always exists under Assumption \ref{a:intersect}.} 
		\[
		F^\star\in \Fdes \text{\; and \;} \im \bbm F^\star \\ I_s \ebm \subseteq \im \bbm X_1 \\ Z_0 \ebm.
		\]
		\item Choose $K\in \mathcal{K}$. 

	\end{enumerate}
	\medskip{}
{Next, we apply the above procedure to a series of illustrative examples that demonstrate its applicability and versatility across different scenarios.}

	\begin{example}[Stabilization via linearization]\label{ex:stab-lin}
	As the first example, we consider the control objective of rendering the equilibrium {of a nonlinear system} asymptotically stable by stabilizing the linearized dynamics. 
	{We note that the data is obtained from the original nonlinear system.}
	It is easy to see that the set of desired closed-systems takes the form:
		\be\label{e:linz}
		\Fdes:=\{F: \exists P>0, \quad FZ'(0)P (FZ'(0))^T-P<0\}, 
		\ee
		where $Z'$ is shorthand notation for the Jacobian of $Z$, {and, for simplicity, the equilibrium is assumed to be at the origin.} 
		Following the second step of the procedure, we look for $F^\star\in \Fdes \cap \Fadm$.  This gives rise to the following set of constraints:
		\vspace{-0.5cm}
		\bse\label{e:Linz}
		\begin{align}
		FZ'(0)P &(FZ'(0))^T-P<0, \quad P>0,\\
		F&=X_1G,\\
		I_s&=Z_0G.
	   \end{align}
	   \ese		
The aforementioned constraints can be transformed into a LMI through {a suitable} change of variables. {Specifically, by defining $Y := GZ'(0)P$ and employing a Schur complement argument, we obtain
\begin{equation}\label{e:LMI-linz}
\bbm
P & X_1Y \\ Y^TX_1^T & P 
\ebm	>0, \quad Z'(0)P-Z_0  Y=0.
\end{equation}
If the LMI \eqref{e:LMI-linz} admits a solution $(P, Y)$, then the constraints in \eqref{e:Linz} are satisfied for $(P, G)$ with 
\[
G= Z_0^+ + (I- Z_0^+Z_0)YP^{-1}{Z'(0)}^\ell,
\] 
as can be verified by direct substitution,
where $Z_0^+:=Z_0^T(Z_0Z_0^T)^{-1}$ and  ${Z'(0)}^\ell$ is any left inverse of $Z'(0)$. A simple choice of this left inverse is obtained by partitioning the library as in \eqref{part} and setting ${Z'(0)}^\ell= \bbm I_n & 0\ebm.$ The resulting controller from the third step of the procedure is then given by $K=U_0G$.} The LMI \eqref{e:LMI-linz}, obtained systematically, provides a {generalization} of the linear stabilization results \cite[Thm. 3]{de2019formulas} {to nonlinear systems. 	{\xqed{$\square$}}}

\end{example}

	\begin{example}[Stabilization via nonlinearity cancellation]\label{ex:canc}
		Let $Z(x)$ {in \eqref{e:model}} be partitioned as 
		\be\label{part}
		Z(x)= \bbm x \\ Q(x)  \ebm.
		\ee
		where $Q:\R^n \rightarrow \R^{s-n}$ contains all the nonlinear functions in the library.
		The goal here is to stabilize the nonlinear system by rendering the closed-loop dynamics linear {Schur stable}. 
		In this case, the set of desired closed-loop systems is given by
		\[
		\Fdes:=\left\{ \bbm \bar F & 0 \ebm : \exists P>0, \quad \bar F^T P \bar F- P < 0\right\},
		\]
		where the partition is consistent with \eqref{part}, i.e. $\bar F\in \R^{n\times n}$. This concludes the first step of the procedure.
		The second step of the procedure is to search for a matrix $F^\star$ belonging to $\Fdes \cap \Fadm$. Clearly, the intersection is given by 
		\[
		\Fdes \cap \Fadm=
		\left \{\bbm \bar F & 0 \ebm \in \Fdes:   \im \left[\begin{array}{@{}ll@{}} \bar F &  \hspace*{-4mm}0  \\ \hline \quad I_s \end{array} \right] \subseteq \im \bbm X_1 \\ Z_0 \ebm,\right \}.
		\]
		This boils down to finding $\bar F, P, G$ satisfying the following constraints:
		\vspace{-0.3cm}
		\begin{align*}
		\bar F^T P \bar F- P &\, {<}\, 0, \quad P>0,\\
		\bbm \bar F & 0 \ebm &= X_1 G,\\
		I_s&= Z_0G.
		\end{align*}
		A solution satisfying the above constraints exists under Assumption \ref{a:intersect}. By Theorem  \ref{t:main}, given any feasible solution $(\bar F, P, G)$ to the above constraints, the controller $u=KZ(x)$ with $K=U_0G$ solves Problem \ref{prob}, namely the closed-loop system becomes a linear Schur stable system as desired. {The result systematically obtained here} coincides with the recent finding in \cite[Thm. 1]{de2023learning}.
			
		{\xqed{$\square$}}
	\end{example}
{
\begin{example}[Diagonal stabilization]\label{ex:diag-stab}%
Consider a system in continuous-time with $Z(x)$ given by
 \[
 Z(x)= \left[ \begin{smallmatrix}\phi_1(x_1) \\  \phi_2(x_2) \\ \ldots \\   \phi_n(x_n) \\[1mm] Q(x) \end{smallmatrix}\right],
 \]
where $\phi_i:\R \rightarrow \R$ is a strictly monotone map for each $i$, $\phi_i(0)=0$, and $Q:\R^n \rightarrow \R^{s-n}$ contains additional nonlinear functions in the library. 
The desired closed-loop systems are given by
\be\label{e:Mphi}
\dot x= M \phi(x),
\ee
where $\phi(x)= \col(\phi_i(x_i))$ and $M\in \R^{n\times n}$ is a diagonally stable matrix, i.e., $M^TD+DM<0$ for some diagonal positive definite matrix $D$. 
The motivation behind diagonal stability is as follows. Consider the Lyapunov function candidate $V(x):= \sum_{i=1}^n d_i \int_0^{x_i} \phi_i(y_i) dy_i$, with $d_i$ being the $i$th diagonal element of $D$. Then, the time derivative of $V$ along the solutions of \eqref{e:Mphi} is given by
$
\dot V(x)= \phi(x)^T DM \phi(x)\leq 0,
$
where the inequality follows from diagonal stability property of $M$. Subsequently, asymptotic stability of the equilibrium follows from the implication $\phi(x)=0 \Rightarrow x=0$. 
Now, consistent with \eqref{e:Mphi}, define
\[
\Fdes:=\left\{ \bbm M & 0 \ebm : \exists D=\diag(d_i)>0, \,  M^T D + DM < 0\right\}.
\]
Following the proposed procedure, we search for an element in 
$\Fdes \cap \Fadm$, which corresponds to finding $G_1,G_2, D$ satisfying
\begin{align*}
M^T D+D M &<0, \quad D=\diag(d_i)>0, \\
\bar M= &X_1G_1, \quad  0=X_1G_2, \quad  
I_s= Z_0\bbm G_1 & G_2 \ebm.
\end{align*}
The corresponding controller is given by $u=U_0(G_1\phi(x)+G_2Q(x))$.
	{\xqed{$\square$}}
\end{example}
}
{We note that in data-driven stabilization procedures, what is utilized is not the specific Lyapunov function itself but rather a parametrized class of candidate functions. The design procedure searches within this class to identify a suitable Lyapunov function that certifies the stability of the closed-loop system.} {Next, we move away from stabilization and discuss an example with a different control objective.} 
	\begin{example}[Nonlinear oscillator design]\label{osc-des}
		Consider the planar system
		\vspace{-\baselineskip}
		\begin{align*}
		x_1^+ &= {x_1} + x_2\\
		x_2^+ &= f(x_1, x_2)+u 
		\end{align*}
		for some $f:\R^2 \rightarrow \R.$
		The goal is to design a state-feedback control such that the resulting closed-loop system behaves like a Van der Pol Oscillator in discrete-time, i.e.,
		\begin{align*}
		\hat x_1^+ &= {\hat x_1} + \hat x_2\\
		\hat x_2^+ &=  \hat x_2+ {{\mu^2}} (\hat x_2- \frac{1}{3}\hat x_2^3-\hat x_1) 
		\end{align*}
		with 
		$\mu$ satisfying the design constraint ${\mu_\ell} \leq \mu \leq \mu_u$, for some given lower and upper bound.
		
		In this case, we partition the library $Z(x)$ as
		\be\label{VP}
		Z(x)= \bbm x_1 \\ x_2 \\ x_2^3 \\[1mm] Q(x) \ebm,
		\ee
		where $Q:\R^2\rightarrow  \R^{s-3}$ contains all functions that can potentially appear in $f(x_1, x_2)$ other than the first three functions in $Z$. {Note that the first three elements in $Z$ capture all the linear and nonlinear functions appearing in the desired closed-loop dynamics. Then, } the set of desired closed-loop systems is given by
		\[
		\Fdes=\left\{ \bbm {1} & 1 & 0 & 0^T_{s-3} \\ -\mu^2 & 1+\mu^2 & -\frac{1}{3} \mu^2 & 0^T_{s-3}  \ebm:  {\mu_\ell} \leq \mu \leq \mu_u,\right\},
		\]
		where the partition is consistent with \eqref{VP}. Now, applying the second step of the procedure results in the following data-based program:
		\begin{align*}
		\text{find}\;& \mu\in \R; g_1, g_2, g_3\in \R^N; G_4\in R^{N\times (s-3)}\\
		\text{s.t.}& \\
		X_1 &\bbm g_1 & g_2 & g_3 \ebm = 
		\bbm {1} & 1 & 0 \\ -\mu^2 & 1+{\mu^2} & -\frac{1}{3} {\mu^2} &  \ebm, \\
		 X_1G_4&=0,  \;\;
		Z_0 \bbm g_1 & g_2 & g_3 & G_4 \ebm =I_s, \; \mu \in [\mu_\ell, \mu_u].
		\end{align*}
		The controller, as indicated in the third step of the procedure, is given by $K= U_0 \bbm g_1 & g_2 & g_3 & G_4 \ebm$.
	\end{example}

\subsection{{On the solvability of Problem \ref{prob} from data}} \label{ss:nec}

{In this subsection, we demonstrate that the condition 
$\Fadm \cap \Fdes\ne \emptyset$, as assumed in Assumption \ref{a:intersect} and with $\Fadm$ defined in \eqref{Fadm}, 
is generally non-conservative when solving Problem \ref{prob} from data.}

First, recall that Problem \ref{prob}  is solvable if and only if $\FadmM \cap \Fdes\ne \emptyset$, where $\FadmM$ is given by \eqref{Fadm-model}. The following result directly follows from Lemma \ref{l:prop}.
	\begin{corollary}\label{c:nec}
		Assume that the matrix $\left[\begin{smallmatrix} Z_0 \\ U_0 \end{smallmatrix}\right]$ has full row rank.
		Then, Problem  \ref{prob} is solvable if and only if Assumption \ref{a:intersect} holds.
	\end{corollary}

	The above result states that for `rich' datasets, Assumption \ref{a:intersect} is both necessary and sufficient for solving Problem \ref{prob}. {{For linear systems, where $Z(x) = x$, this richness property, specifically the full row rank condition of $[\begin{smallmatrix} Z_0 \\ U_0 \end{smallmatrix}]$, can be guaranteed by ensuring that the input is persistently exciting of order $n+1$,  as established by Willems' fundamental lemma \cite{willems2005note}. For general nonlinear systems, to the best of our knowledge, a full extension of the fundamental lemma is not yet available. Nevertheless, tailored formulations have been developed for specific classes of nonlinear systems. We refer to \cite[Sec. 4.1]{berberich2024overview} for an overview of such results.}

	Next, we argue that in case the rank condition in Corollary \ref{c:nec} does not hold, 
	the condition in Assumption \ref{a:intersect} is still generally ``necessary" for solving Problem \ref{prob} using only system data in \eqref{data}. This is true providing that the closed-loop property  characterized by $\Fdes$ is  ``binding". This property is rather mild as formalized below:
	
	\begin{definition}\label{d:bind}
		We call $\Fdes$  {\em binding} if the following implication holds:
		\[
		F^\star\in \Fdes, \; v\in \R^{s}\setminus \{0_s\} \Longrightarrow \exists w\in \R^{n} \; {\rm s.t.} \; F^\star+ wv^\top \notin \Fdes.
		\]
	\end{definition}
	
	\medskip{}
	Note that that the binding property is independent of the data set.
	The property requires that for any $F^\star\in \Fdes$ and nonzero vector $v$, there exists a vector $w$ such that the perturbed matrix $F^\star+wv^T$ leaves the desired set $\Fdes$.  The following lemma provides {two} notable special cases for binding desired sets.  
	\begin{lemma}\label{l:binding}
		The set $\Fdes$ is binding if either of the following conditions hold:
		\begin{enumerate}[(i)]
			\item $\Fdes$ is finite.
			\item $\Fdes$ is bounded in some matrix norm.
		\end{enumerate}	
	\end{lemma}

	\BP
	The first condition is trivial in view of Definition \ref{d:bind}. For the second condition, we have
	\[
	\norm{F^\star+wv^T}_F^2= \norm{F^\star}_F^2+ 2\underbrace{w^TF^\star v}_{\geq -\norm{w}\norm{F^\star v}} + \norm{w}_2^2 \norm{v}_2^2.
	\] 
	Therefore, choosing $\norm{w}$ arbitrary large makes $F^\star+wv^T$ unbounded in Frobenius norm and thus in any {matrix norm}. Hence, {for any $v$}, there always exists $w$ such that, $F^\star+wv^T\notin \Fdes$ implying that $\Fdes$ is binding
	\EP

	The first condition of Lemma \ref{l:binding}  states that any $\Fdes$ with a {\em{finite}} cardinality is binding. This  captures the scenarios where the desired  controller coincides with the unique solution (even locally) of an optimization problem as well as the case of prescribing a desired closed-loop system.   {The second condition states that $\Fdes$ is binding if the norm of the matrices in $\Fdes$ admit a uniform bound, i.e. there exists $\alpha$ such that $\norm{F^\star}\leq \alpha$, $\forall F^\star\in \Fdes.$}

	{An example of a property that is not generally binding is the case where the only requirement of $\Fdes$ is asymptotic stability of the equilibrium. In particular, consider the set $\Fdes$ as given by \eqref{e:linz} and let 
	$Z(\cdot)$ be partitioned as \eqref{part} with $s>n$.
	Consistently, let $v=\bbm v_1^T & v_2^T\ebm^T$ and choose $v_1^T=-v_2^TQ'(0)$, where $Q'$ is the shorthand notation for the Jacobian of  $Q$. Clearly, by construction $v^TZ'(0)=0$. 
Therefore, noting \eqref{e:linz}, for any $F^\star\in \Fdes$, the perturbed matrix  $F^\star+wv^T$ remains in $\Fdes$ for all $w\in \R^n$, which implies that $\Fdes$ is not binding.

Interestingly, the asymptotic stability property becomes binding if some guarantees on the region of attraction are also required in $\Fdes$. To see this, suppose that the desired closed-loop systems are given by those whose equilibrium is asymptotically stable, and there exists a {compact} estimate of region of attraction\footnote{An estimate of region of attraction is a forward invariant set such that any solution initialized in this set asymptotically converges to the origin.} $\Omega$ with $\mathcal{B}_{\delta_1}\subseteq \Omega \subseteq \mathcal{B}_{\delta_2}$ for some prescribed $\delta_1, \delta_2>0$. {Here, $\mathcal{B}_\delta:=\{\norm{x}\leq \delta\}$ denotes the ball of radius $\delta$ centered at the origin,}
the lower bound is to exclude arbitrary small estimates of region of attraction, and the upper bound enforces compactness of $\Omega$.
To investigate if this property is binding, given $v\in \R^s$, consider the family of systems
\be\label{e:family}
x^+= (F^\star+wv^T)Z(x), \quad w\in \R^n,
\ee
and suppose $F^\star\in \Fdes$.
Let
\[
x_0:={\rm{argmax}}_{x\in \mathcal{B}_{\delta_1}}\, |v^TZ(x)|.
\]
The value of $|v^TZ(x_0)|$ is {finite and} nonzero under the mild assumption that the functions in $Z(\cdot)$ are linearly independent.
Initializing \eqref{e:family} at $x(0)=x_0$, we have
\begin{align*}
	\norm{x(1)}_2^2&=\norm{ (F^\star+ wv^T)Z(x_0)}^2\\ 
	&=  Z(x_0)^T{F^\star}^TF^\star Z(x_0)+ 2Z(x_0)^T{F^\star}^T wv^T Z(x_0)\\
	&\qquad +\norm{w}^2 \norm{v^T Z(x_0)}^2.
\end{align*}
Clearly, as $w$ becomes arbitrarily large, $x(1)$ leaves any  given compact set $\Omega\subseteq \mathcal{B}_{\delta_2}$, and thus $\Fdes$ is binding in this case.}

	The following result shows that, for binding properties, Assumption \ref{a:intersect} is ``necessary" for solving Problem \ref{prob} using only data.\footnote{Such notion of ``necessity" is formalized and discussed for various linear control problems in \cite{van2020data}.}

	\begin{theorem}
		Assume that $\Fdes$ is binding and that $\Fadm \cap \Fdes = \emptyset$.\footnote{The empty intersection means that Assumption \ref{a:intersect} does not hold.} Suppose that there exists a feedback $u=KZ(x)$ solving Problem \ref{prob}, i.e., $A+BK\in \Fdes$. Then, there exist $\bar A$ and $\bar B$ consistent with the data, namely $X_1=\bar A Z_0 + \bar B U_0$ such that $\bar A+ \bar B K \notin \Fdes.$
	\end{theorem}

	\BP
	By the hypotheses of the theorem, we find that $A+BK\notin \Fadm$, i.e,
	$
	\im \left[ \begin{smallmatrix} A+BK \\ I_s \end{smallmatrix}\right]  \nsubseteq \im \left[ \begin{smallmatrix} X_1 \\ Z_0 \end{smallmatrix}\right] .
	$
	Hence, 
	\[
	\im \bbm A & B \\  I_s & 0 \ebm \bbm I_s \\ K \ebm \nsubseteq \im \bbm A & B \\  I_s & 0 \ebm \bbm Z_0 \\ U_0 \ebm,
	\]
	and thus
	\[
	\im \bbm I_s \\ K \ebm \nsubseteq \im\bbm Z_0 \\ U_0 \ebm,
	\]
	or equivalently
	\[
	\ker \bbm Z_0 \\ U_0 \ebm^T  \nsubseteq  \ker  \bbm I_s \\ K \ebm^T.
	\]
	Therefore, there  exists a nonzero vector 
	$\tilde v=\col(\tilde v_A, \tilde v_B)$, such that
	\be\label{ker-rel}
	\bbm \tilde v_A^T& \tilde v_B^T\ebm  \bbm Z_0 \\  U_0\ebm=0, \quad  \bbm\tilde v_A^T& \tilde v_B^T\ebm  \bbm I_s\\  K\ebm\ne 0.
	\ee
	Due to the first equality, it is easy to see that the matrices $\bar A:=A+w\tilde v_A^T$ and $\bar B:=B+w\tilde v_B^T$ satisfies the equality  $X_1=\bar A Z_0 + \bar B U_0$ for all $w\in \R^n$.
	Define
	\[
	v^T:= 
	\bbm \tilde v_A^T& \tilde v_B^T\ebm \bbm I_s\\ K\ebm. 
	\]
	The vector $v$ is nonzero due to \eqref{ker-rel}.
	For any $w\in \R^n$, we have
	$
	\bar A+ \bar B K= (A+BK) + wv^T.
	$
	Then, bearing in mind that $A+BK\in \Fdes$ and $\Fdes$ is binding, there exists  $w\in \R^n$ such that 
	$\bar A+ \bar B K \notin \Fdes,$ which completes the proof. 
	\EP

	\vspace{-2mm}	
	\subsection{Extension to open systems}\label{ss:open}
	In this subsection, we show how the results can be extended to {feedforward-feedback} control law
	$u=K Z(x)+ K_r r$, with control gains $K \in \R^{m\times s}$, $K_r \in \R^{m\times m_r}$ and an external  input $r\in \R^{m_r}$ with $m_r \leq m$.  This results in the open system 
	\be\label{e:open}
	x^+=(A+BK)Z(x)+BK_r r
	\ee
	The set of desired controlled system can be then parametrized as (cf. \eqref{CLdes})
	\be\label{des-open}
	x^+= \bbm F^\star  & F^\star_r \ebm \bbm Z(x)\\ r \ebm, \quad    \bbm F^\star  & F^\star_r \ebm \in \Fdes,
	\ee
	where $\Fdes$ now dictates the desired specifications for both matrices $F\equiv A+BK$ and $F_r\equiv BK_r$ of the controlled system \eqref{e:open}.

	\begin{problem}\label{prob-open}
		Given $\Fdes$, design a control law $u=KZ(x)+K_r r$ such that the resulting controlled system  satisfies \eqref{des-open}. Equivalently, find $K$ and $K_r$ such that  $\bbm A+BK & BK_r\ebm \in \Fdes$.
	\end{problem}

	By \eqref{e:open}, the set of attainable controlled systems is obtained as (cf. \eqref{Fadm-model})
	\be\label{Fadm-model-open}
	\FadmM:= \{\bbm F  & F_r \ebm :  \im \bbm A-F  & F_r \ebm \subseteq \im B \}.
	\ee
	Next we separate the system matrices  from $F$ and $F_r$ in the above subspace inclusion. It is easy to verify that we can equivalently write the set in  \eqref{Fadm-model-open} as (cf. \eqref{nec-1})
	\be\label{admM-open}
	\FadmM=\left \{\bbm F  & F_r \ebm: \im \bbm F & F_r \\ I_s  & 0 \ebm \subseteq \im {\bbm A & B \\I_s & 0 \ebm} \right\}. 
	\ee
	With the same principle as before, since the system matrices are not available, we work with a purely data-based subspace of $\im \bbm\begin{smallmatrix} A & B \\I_s & 0\end{smallmatrix} \ebm$. This subspace is given by the left hand side of  \eqref{nec-2}.
	By substituting this data-based subspace in place of $\im \bbm\begin{smallmatrix} A & B \\ I_s & 0\end{smallmatrix} \ebm$ in \eqref{admM-open}, we obtain the set (cf. \eqref{Fadm})
	\be\label{Fadm-open}
	\Fadm:=\left \{\bbm F  & F_r \ebm: \im \bbm F & F_r \\ I_s  & 0 \ebm  \subseteq \im \bbm X_1 \\ Z_0 \ebm \right \}.
	\ee
	All statements of Lemma \ref{l:prop} holds for the set $\Fadm$ given by \eqref{Fadm-open}.
	Indeed, the inclusion $\Fadm\subseteq \FadmM$ holds by noting \eqref{nec-2}, \eqref{admM-open}, and \eqref{Fadm-open}. The proof of the other two statements are analogous to the arguments provided in the proof of the lemma. Now, the counterpart of Theorem \ref{t:main} for open systems is provided below:
	
	\begin{theorem}\label{c:main}
		Let Assumption \ref{a:intersect} hold with $\Fdes$ the set of desired controlled system in \eqref{des-open}  and $\Fadm$ given by \eqref{Fadm-open}. 
		Define
		\vspace{-0.3cm}
		\begin{align}\label{calK-open}
		\mathcal{K}_{\rm ext}:=\Big\{ \bbm K & K_r\ebm:& \im \bbm F^\star & F_r^\star \\ I_s & 0 \\ K & K_r  \ebm  \subseteq \im \bbm X_1 \\ Z_0 \\ U_0 \ebm,\nonumber  \\ 
		&\qquad {\bbm F^\star & F^\star_r\ebm \in \Fdes} \Big\}. 
		\end{align}
		Then,  the set $\mathcal{K}_{\rm ext}$ is nonempty. Moreover, Problem \ref{prob-open} is solvable by the control law $u=KZ(x)+K_r r$ for any $\bbm K & K_r\ebm\in \mathcal{K}_{\rm ext}$. 
	\end{theorem}
	
	\BP
	{The fact that the set $\calK$ is nonempty follows from the assumption and \eqref{Fadm-open}.
Choose $\bbm K & K_r\ebm \in \mathcal{K}_{\rm ext}$.
The subspace inclusion in \eqref{calK-open} can be split to
\[
\im \bbm F^\star \\ I_s \\ K \ebm  \subseteq \im \bbm X_1 \\ Z_0 \\ U_0 \ebm, 
\quad \im \bbm F_r^\star \\ 0 \\ K_r  \ebm  \subseteq \im \bbm X_1 \\ Z_0 \\ U_0 \ebm.
\]
Analogously to the proof of Theorem \ref{t:main}, the first subspace inclusion implies that $A+BK=F^\star$.
By the second subspace inclusion, there exists a matrix $G_r\in \R^{N\times m_r}$ such that $F_r^\star=X_1G_r$, $0=Z_0G_r$ and $K_r=U_0G_r$. Then, from \eqref{data}, it follows that $BK_r=F_r^\star$.
The proof is complete noting that $\bbm F^\star & F^\star_r\ebm \in \Fdes$.}
	\EP

	Theorem \ref{c:main} provides a  {systematic} procedure for solving Problem \ref{prob-open} using data:
	\begin{enumerate}
		\item Specify the set of desired controlled system $\Fdes$ based on the design objective.
		\smallskip{}
		\item Find $\bbm F^\star & F_r^\star\ebm \in \Fdes \cap \Fadm$. Namely, find $F^\star$ and $F_r^\star$ such that
		\vspace{-3mm}
		\[
		\bbm F^\star & F_r^\star\ebm \in \Fdes \text{\; and \;} \im \bbm F^\star & F_r^\star 
		\\ I_s & 0\ebm \subseteq \im \bbm X_1 \\ Z_0 \ebm.
		\]
		\vspace{-3mm}
		\item Choose {$\bbm K & K_r\ebm \in \mathcal{K}_{\rm ext}$.} 
	\end{enumerate}

	\begin{example}[{Nonlinear} model reference control]\label{e:MRC}
		Consider a reference nonlinear model
		\be\label{e:ref}
		x_{\rm ref}^+=\bar A \bar Z (x_{\rm ref})+ \bar Br,
		\ee
		with state $x_{\rm ref}\in \R^n$, input $r\in \R^{m_r}$, $\bar A \in \R^{n \times \bar s}$, and $\bar Z: \R^n \rightarrow  \R^{\bar s}$ capturing the nonlinearities appearing in the reference model.
		The model reference control (MRC) problem that we consider here is to find the matrices $K$ and $K_r$ such that the 
		following matching conditions hold:
		\[
		(A+BK)Z(x)=\bar A \bar Z(x) , \quad BK_r= \bar B, \quad \forall x\in {\R^n},
		\]
		Now, assume that  the functions in $\bar Z$ is a subset of those in $Z$\,\footnote{If this is not the case, the library $Z(\cdot)$ can be simply extended to include any additional functions appearing in the reference model.}  and partition $Z$ as $Z(x)=\col(\bar Z(x), \tilde{Z}(x))$, with $\tilde{Z}: \R^n \rightarrow  \R^{s-\bar s}$.
		Then, the MRC problem can be equivalently recast as finding  the matrices $K$ and $K_r$ such that
		\[
		A+BK= \bbm \bar A & 0 \ebm, \quad BK_r= \bar B.
		\]
		Hence, the set of desired controlled systems in \eqref{des-open} is specified by the singleton $\Fdes=\{\bbm F^\star & F_r^\star\ebm\}$, with $F^\star:= \bbm \bar A & 0 \ebm$ and  $F_r^\star:=\bar B.$
		Next, following the second step of the procedure, the following program solves the nonlinear MRC problem:
		\begin{align*}
		\text{find}\;& G_{11}\in \R^{N\times \bar s}, G_{12}\in \R^{N\times (s-\bar s)}, G_2\in \R^{N\times m_r}\\
		\text{s.t.}& \\[-2mm]
		&\bbm \bar A & 0   &  \bar B \\ I_{\bar s} & 0 & 0\\ 0 & I_{s-\bar s} & 0 \ebm = \bbm 
		X_1 \\ \bar Z_0 \\  \tilde Z_0\ebm \bbm G_{11} & G_{12} & G_2 \ebm,
		\end{align*}
		where the data matrix $Z_0$ is partitioned as $\col(\bar Z_0, \tilde Z_0)$ consistent to the partitioning of $Z$. 
		The resulting control law is then given by the third step of the procedure as $$u=KZ(x)+K_rr=  U_0G_{11}\bar Z(x)+ U_0G_{12}\tilde Z(x)+ U_0G_2r.$$
		In the special case where the functions in $Z(\cdot)$ coincides with those in the reference model, the control law reduces to $u=U_0G_{11}\bar Z(x)+ U_0G_2r.$ A notable example of this special case is MRC in linear systems where $Z(x)=\bar Z(x)=x$.  For linear systems, the resulting controller gains coincide with those reported in \cite{breschi2021direct,wang2023necessary,padoan2023controller}.  	{\xqed{$\square$}}
	\end{example}
	
	\begin{example}[Feedback cyclo-passivation]
		Consider the system \eqref{e:open} in continuous-time. The problem of interest is to design the controller gains $K$ and $K_r$ such that the system becomes passive from the input $v$ to a suitably defined output $y:=h(x)$, $h:\R^n \rightarrow \R^{m_r}.$ Namely, there should exist a storage function $S:\R^n \rightarrow  \R$ such that\footnote{We use a variation of passivity, sometimes referred to as cyclo-passivity, where nonnegativity of the storage function is not required.} 
		\[
		\frac{\partial S}{\partial x}^T(x) (A+BK)Z(x)+\frac{\partial S}{\partial x}^T BK_r r\leq r^T h(x),  \quad \forall x, r.
		\]
		This dissipation inequality holds if and only if \cite[Prop. 4. 1. 2]{van2016l2},
		\vspace{-2mm}
		\bse\label{e:pass}
		\be\label {pass-1}
		\frac{\partial S}{\partial x}^T(x) (A+BK)Z(x)\leq 0
		\ee
		\vspace{-2mm}
		\be\label{pass-2}
		\frac{\partial S}{\partial x}^T BK_r=h^T(x).
		\ee
		\ese
		Suppose that the gradient of the storage function can be written as a linear combination of the functions in the library, i.e. $\frac{\partial S}{\partial x}=MZ(x)$ fo some matrix $M\in \R^{n \times s}.$ Similarly, we write  the output as $h(x)=NZ(x)$ with $N\in \R^{m_r\times s}.$ 
		The conditions \eqref{e:pass} are then satisfied if
		\be\label {pass-3}
		Z^T(x)M^T (A+BK)Z(x)\leq 0
		\ee
		and 
		\be\label{pass-4}
		M^TBK_r= N^T.
		\ee
		Let 
		\be\label{classM}
		\mathcal{M}:=\{M\mid M\frac{\partial Z}{\partial x}=\frac{\partial Z}{\partial x}^TM^T, \quad \forall x\}.
		\ee
		Then the set of desired controlled system is specified in terms of $F\equiv A+BK$ and $F_r \equiv BK_r$ as   
		\begin{align}\label{fdes-passive}
		\Fdes:=&\{\bbm F & F_r\ebm: Z(x)^TM^T F Z(x) \leq0,   \nonumber \\
		&\qquad M^TF_r =N^T, \; M\in \mathcal{M}, \; N \in \R^{m_r\times s}\}. 
		\end{align}
		{Note that the inequality in \eqref{fdes-passive} corresponds to \eqref{pass-3}, the equality corresponds to \eqref{pass-4}, and the constraint $M\in \mathcal{M}$ is included to ensure that $MZ(x)$ can be written as a gradient of a function $S(\cdot)$.}
		Now, Theorem \ref{c:main} yields the following data-dependent program:
		\begin{align}\label{pass-org}
		\nonumber
		\text{find}\; &  F^\star\in \R^{n\times s}, F^\star_r  \in \R^{n\times m_r}, K\in \R^{m\times s}, K_r \in \R^{m\times m_r}\\
		\nonumber
		\text{s.t.}& \\[-3mm]
		&\im \bbm F^\star & F_r^\star \\ I_s & 0 \\ K & K_r  \ebm  \subseteq \im \bbm X_1 \\ Z_0 \\ U_0 \ebm,
		\;
		\bbm F^\star & F^\star_r  \ebm \in \Fdes 
		\end{align}
	with $\Fdes$ given by \eqref{fdes-passive}.
		Note that the above program does not contain any model information apart from the library $Z(\cdot)$.
		By tweaking the above program and putting in some additional effort, we obtain the following result as a special case.
	\end{example}
	
	\begin{corollary}\label{c:passive}
		Let $M\in \mathcal{M}$.  Suppose that there exist $G_1\in \R^{N\times s}$, $G_2\in \R^{N\times m_r}$, and $\Theta\in \R^{n\times n}$ such that
		\bse\label{pass-BMI}
		\begin{align}
		\label{pass-g1g2}
		\bbm I_s & 0 \ebm &= Z_0 \bbm G_1 & G_2 \ebm.\\
		\label{pass-theta1}
		X_1G_1-\Theta M&=0,\\
		\label{pass-theta2}
		\Theta+ \Theta^T &\leq 0.
		\end{align}
		\ese
		Then the control law $u=KZ(x)+K_rr$ with $K=U_0G_1$ and $K_r=U_0G_2$, renders the controlled system \eqref{e:open} passive from input $r$ to output $y:=G_2^TX_1^T MZ(x)$. Moreover, a storage function certifying passivity is given by 
		\be\label{storage}
		S(x)=\int_0^1 x^T MZ(tx) dt. 
		\ee
	\end{corollary}
	
	\begin{remark}
		We briefly comment on the feasibility of \eqref{pass-BMI}.
		The equality constraint \eqref{pass-g1g2} is feasible if and only if $Z_0$ has full row rank. 
		The equality constraint \eqref{pass-theta1} holds for some $\Theta$ if and only if $\ker M \subseteq \ker X_1G_1$.
		Overall, for a given $M$, the constraints form an LMI in variables $G_1$ and $\Theta$, while $G_2$ is only used to shape the passive output. Hence, one needs to search for a matrix $M$ in class $\mathcal{M}$ such that the corresponding LMI in \eqref{pass-BMI} is feasible. 
	\end{remark}
	
	We note that after applying the control law in Corollary \ref{c:passive}, the resulting controlled system \eqref{e:open} takes the form
	\begin{align*}
	x^+&= X_1G_1Z(x)+ X_1G_2 r = \Theta MZ(x) + X_1G_2r,\\ y&= G_2^TX_1^T MZ(x).
	\end{align*}
	Noting that $M\in \mathcal{M}$, we can write $MZ(x)=\frac{\partial S}{\partial x}$ for some function $S:\R^n \rightarrow \R$. Hence, the controlled system simplifies to
	\[
	x^+= \Theta \frac{\partial S}{\partial x}+ X_1G_2 r, \quad y= (X_1G_2)^T\, \frac{\partial S}{\partial x}.
	\]
	Using \eqref{pass-theta2}, it turns out the above controlled system admits a port-Hamiltonian representation with $S$ serving as the Hamiltonian \cite[Ch. 6]{van2016l2}. The fact that port-Hamiltonian systems are passive proves the claim made in Corollary \ref{c:passive}; namely, 
	\[
	\frac{\partial S}{\partial x}^Tx^+= \frac{\partial S}{\partial x}^T \Theta \frac{\partial S}{\partial x}+\frac{\partial S}{\partial x}^T X_1 G_2r \leq   \frac{\partial S}{\partial x}^T X_1 G_2r =r^T y,
	\]
	where we used \eqref{pass-theta2} to write the inequality. The explicit form of the storage function in \eqref{storage} follows from the Hadamard lemma \cite[Ch. 2]{nestruev2003smooth}, noting $\frac{\partial S}{\partial x}=MZ(x)$.
	It is worth mentioning that a systematic way for feedback passivation of nonlinear systems  is in general missing even in the model based regime. Specialising the class of passive systems to port-Hamiltonian systems allows to write the more tractable conditions \eqref{pass-BMI} in place of \eqref{pass-org}. For linear systems, the two classes coincide.  In particular, restricting the results to linear systems $Z(x)=x$ and positive definite storage functions, the class \eqref{classM} modifies to positive definite $n\times n$ matrices and the constraints in \eqref{pass-BMI} reduce to 
	\vspace{-\baselineskip}
	\begin{align}\label{e:LMI-p-lin}
	\bbm I_s & 0 \ebm &= Z_0 \bbm G_1 & G_2 \ebm, \nonumber \\
	X_1G_1M^{-1} &+M^{-1}(X_1G_1)^T\leq 0, \quad M>0.
	\end{align}
	It is easy to show that the above inequalities can be equivalently stated as the following linear matrix inequalities 
	\[
	0=Z_0G_2, 
	\quad
	X_1Q +(X_1Q)^T\leq 0, \quad 
	Z_0Q=(Z_0Q)^T>0,
	\]
with variables $Q$ and $G_2$. 
If the above LMI is feasible, then \eqref{e:LMI-p-lin} is satisfied with $M:=(Z_0Q)^{-1}$ and $G_1:=Q(Z_0Q)^{-1}$, and
the controller $u=U_0G_1x+ U_0G_2r$ renders the linear system passive from input $r$ to output $y=G_2^TX_1^T (Z_0Q)^{-1}x$. The underlying storage function in \eqref{storage} takes the quadratic form 
\vspace{-\baselineskip} 
\[
S(x)={\int_0^1 t\,x^TMx \,dt=} \frac{1}{2}x^TMx=\frac{1}{2} x^T (Z_0Q)^{-1}x.
\]
	
\vspace*{-0.3cm}
\section{Conclusions}\label{s:conc}
We have introduced a unifying and versatile framework for direct data-driven control of nonlinear systems. This framework addresses the sparsity observed in the literature and put forward systematic procedures for synthesizing control algorithms directly from data. 
{By developing a systematic framework, we have shown how several existing results can be incorporated and extended, enabling the development of new results in the domain, particularly in scenarios like nonlinear stabilization, nonlinear oscillator design, model reference control, and passivating feedback.} {Remarkably, regardless of the specific instance of the control problem, the underlying data-driven design principle reduces to three steps: i) specify the set of desired closed-loop systems ($\Fdes$), ii) intersect this set with the set of closed-loop systems attainable by data ($\Fadm$), and iii) select an element from this intersection and choose the corresponding controller gain.}
The examples discussed demonstrate not only the applicability of the proposed framework, but also its potential to foster further advancements in data-driven control research. As the challenges of nonlinear systems and the data they generate continue to grow, the need for effective data-driven control solutions becomes increasingly critical. While the discussed framework is well-positioned to meet these challenges, there is still a long way to go in providing complete answers to the complexities inherent in nonlinear data-driven control. 
{A particularly relevant extension of this work would be to address the neglected nonlinearities in the considered library of functions}
\bibliographystyle{IEEEtran}
\bibliography{collection}
\end{document}